\documentstyle[11pt,aasms4]{article}
\begin{document}
\baselineskip 18.0pt
\def\oneskip{\vskip\baselineskip}
\def\xr#1{\parindent=0.0cm\hangindent=1cm\hangafter=1\indent#1\par}
\def\la{\raise.5ex\hbox{$<$}\kern-.8em\lower 1mm\hbox{$\sim$}}
\def\ma{\raise.5ex\hbox{$>$}\kern-.8em\lower 1mm\hbox{$\sim$}}
\def\ea{\it et al. \rm}
\def\am{$^{\prime}$\ }
\def\as{$^{\prime\prime}$\ }
\def\msol{M$_{\odot}$ }
\def\kms{$\rm km\, s^{-1}$}
\def\cm3{$\rm cm^{-3}$}
\def\Ts{$\rm T_{*}$}
\def\Vs{$\rm V_{s}$}
\def\n0{$\rm n_{0}$}
\def\B0{$\rm B_{0}$}
\def\Fh{$\rm F_{H}$}
\def\Fn{$\rm F_{\nu}$}
\def\ne{$\rm n_{e}$}
\def\Te{$\rm T_{e}$}
\def\Hb{H$\beta$}
\def\Tgr{$\rm T_{gr}$}
\def\Tgas{$\rm T_{gas}$}
\def\Ec{$\rm E_{c}$}
\def\N{$\it N$}
\def\ff{$\it ff$}
\def\sf{$\it \sqrt{ff}$}
\def\erg{$\rm erg\, cm^{-2}\, s^{-1}$}
\def\ROIII{$\rm R_{OIII}$}

{\centerline{\Large{\bf The optical - ultraviolet 
continuum of Seyfert 2 galaxies}}}

\bigskip

\bigskip

\bigskip

\centerline{ $\rm M. \, Contini^1 \,\, and \,\,\, S.M. \, Viegas^2  $}

\bigskip

\bigskip

\bigskip

$^1$ School of Physics and Astronomy, Tel-Aviv University, Ramat-Aviv, Tel-Aviv,
69978, Israel

$^2$ Istituto Astron\^{o}mico e Geof\'{i}sico, USP, Av. Miguel Stefano, 
4200,04301-904
S\~{a}o Paulo, Brazil

\bigskip

\bigskip

\bigskip

\bigskip

\bigskip

\bigskip

\bigskip

Running title : The continuum of Seyfert 2 galaxies

\bigskip

\bigskip

\bigskip

\bigskip

\bigskip

\bigskip

subject headings : galaxies : Seyfert - galaxies:  - continuum -
galaxies: shock waves - ultraviolet: galaxies

\newpage

\section*{Abstract}

This paper  aims to  understand
the continuum of Seyfert 2  galaxies.
By fitting the single galaxies in the sample of
Heckman et al. (1995) with composite models (shock+
photoionization),  we show that five main components
characterize the SED of the continuum.
Emission in the radio range  can be recognized as bremsstrahlung
 from relatively cold gas,
or synchrotron radiation due to Fermi mechanism at  the shock front.
The bump in the IR  is  due to reradiation of the central radiation by 
 dust and to mutual heating and cooling between dust and gas.
In the optical-UV range the main component is due to
bremsstrahlung from  gas heated and ionized by the primary flux 
(the flux from the active center (AC)  or/and the radiation from young stars),
by diffuse  radiation emitted by the hot slabs of gas,
  and by collisional ionization.
Shocks  play an important role since  they produce a high temperature
zone where soft X-rays are emitted. Finally,  
 the harder X-ray  
radiation  from the AC, seen through the clouds, is easily recognizable
in the spectrum. Assuming that the NLR is powered by a power-law
central ionizing radiation and by shocks, we discuss  the
optical-ultraviolet featureless continuum of Seyfert 2. We show that
in this wavelength range, the slope of the NLR emission reproduces
the observed values, and may be the main component of the featureless
continuum. 
However, the presence of star forming regions  cannot be excluded 
in the circumnuclear region of various Seyfert galaxies. Their photoionizing
radiation may 
prevail in the outskirts
of the galaxy where the power-law radiation from the AC is diluted.
An attempt is made to find their fingerprints in the observed AGN
spectra. Finally, it is demonstrated that multi-cloud 
models are necessary to interpret the spectra of single objects, 
even in the global investigation of a sample of galaxies.

\newpage

\section{Introduction}

Observational data in the ultraviolet range are now available for more and more
galaxies. They complete the range of the observed frequencies 
and permit a better interpretation of the emitted spectra.
The nature of the ultraviolet continuum in type 2 Seyfert galaxies
was investigated by Heckman et al. (1995) on the basis of the IUE spectra
of 20 galaxies (hereinafter, this sample is referred
to as HS). Various possibilities were considered to explain
the observed featureless continuum, as, for example, 
light from a hidden Seyfert 1 nucleus
scattered by dust or warm electrons.  The results show that  no more 
than 20 \%  of the Seyfert 2 template's continuum
can be light from a hidden Seyfert 1 nucleus. The alternative favored by 
Heckman et al.
is that most of the UV continuum in these galaxies 
is produced by a reddened circumnuclear starburst.
 Heckman et al. claim that the UV spectral
slopes and the ratios of far IR to UV continuum fluxes are very similar
to the corresponding properties of typical metal-rich, dusty starburst galaxies.

The interpretation of the observed  continuum 
(and line) spectra of the galaxies is difficult due to the complex
structure of the emitting regions. Modeling is crucial to disentangle 
the different components contributing to a single galaxy spectrum.

Previous papers on active galactic nuclei (AGN) and starburst galaxy emission 
spectra (Contini, Prieto, \& Viegas 1998a,b, Viegas, Contini, \& Contini
1999, Contini et al 1999) have shown that
multi-cloud models are necessary to fit the observational data  even in a  
single region of a galaxy. 
Moreover,  the spectral energy distribution  (SED) of  
AGN and starburst galaxy continua can be 
 roughly decomposed into  five  components.
Emission in the radio range  can be recognized as bremsstrahlung
 from relatively cold gas,
or synchrotron radiation due to the Fermi mechanism at the shock front.
The bump in the IR  is  due to reradiation by dust  of the central 
radiation. 
In the optical-UV range the main component is due to 
bremsstrahlung from  the clouds photoionized by
the flux from the active center (AC) and the radiation from young stars,
by diffuse  radiation emitted by the hot slabs of gas,
as well as  by collisional ionization.
The shock  plays an important role since it produces a high temperature
zone emitting soft X-rays . Finally,  
 the hard X-ray  
radiation  from the AC, seen through the clouds or through the
dust torus, is easily recognizable
in the spectrum.
Previous results (Contini, Prieto \& Viegas 1998a,b) show that 
the SED of the continuum
in the various frequency ranges are often correlated.

In order to further investigate  the nature of the continuum of HS galaxies,
 we have simulated the galaxy spectra by multi-cloud
composite models. These models account for the photoionizing effect
of the radiation from the AC
source, as well as  for  shock effects on the emitting clouds. 

One of our goals is to  search  for  
starburst or  AGN characteristics  prevailing
in single objects and see how these can be recognized by the analysis of the
spectra.
Actually, we focus on the continuum, looking for 
the features that are easily 
recognized in the observed spectra. 
We show that for consistency, however,  both the continuum and line spectra 
must be considered. 
The physical conditions  in single galaxies and in the 
whole sample are also investigated.  
Another goal is to verify if the continuum obtained from the components
listed above reproduces the observed characteristics of the
HS galaxies, in particular
the slopes discussed by Tran (1995) and by Heckman et al. (1995).

For all the objects of the sample, the fit of the observed continuum SED
  and the main characteristics
of the models explaining the observed continuum are presented in \S 2. 
The line ratios are
presented and discussed in \S 3. Concluding remarks follow in \S 4.

\section{The SED of the continuum}

We consider composite models for the NLR 
which account consistently for the effects of an ionizing radiation flux 
from an external source and of the shocks due to cloud  motions.
The SUMA code (see, for instance, Viegas \& Contini 1994) is used.

The input parameters are the shock velocity, \Vs, the preshock density,
\n0, the preshock magnetic field, \B0, the ionizing radiation spectrum,
the chemical abundances, 
the dust-to gas ratio by number, d/g,
and the geometrical thickness of the clouds, D.  A power-law, 
characterized by the power index $\alpha$ and the 
flux, \Fh, at the Lyman limit, reaching the cloud (in units of cm$^{-2}$
s$^{-1}$ eV$^{-1}$) is generally adopted. However, for some models
a high temperature (\Ts=1-2 $10^5$ K) blackbody ionizing radiation is used, 
characterized by the 
ionization parameter U. This high blackbody temperature  could be
associated to an evolved stellar cluster (Terlevich \& Melnick 1985,
Cid-Fernandes, Dottori, Gruenwald \& Viegas 1991). Some models with a blackbody
spectrum at lower temperature (\Ts=5 $10^4$ K) are also considered,
in order to mimic ionization by a starburst. 

For all the models,  \B0 = $10^{-4}$ gauss, $\alpha$ = 1.5, 
and cosmic abundances (Allen 1973) are adopted. The remaining input 
parameters
are listed in Table 1. Notice that some of the models differ only  in
the dust-to-gas ratio.
 SUMA accounts  for silicate grains
with an initial radius of 0.2 $\mu$m. 
The grains are sputtered
entering the shock front (see Viegas \& Contini 1994).
The d/g ratio by mass in the Galaxy is $\sim$ 4.1 $10^{-4}$
(Drain \& Lee 1994) which corresponds to d/g $\sim 10^{-14}$  by number,
adopting a silicate density of $\sim$3 g \cm3.

Previous results, obtained by self-consistently fitting the
continuum  and  emission-line spectra of the Circinus galaxy and NGC 5252,
confirm that velocities of 100-1000 \kms ~are present in the
narrow line region of Seyfert 2 galaxies, with preshock densities of 100-1000
\cm3. The ionizing flux from the AC are about $10^{11} - 10^{12}$ photons $\rm
cm^{-2} \, s^{-1} \, eV^{-1}$ at 1 Ryd.

\subsection{Theoretical results : the optical-UV peak}

The calculated optical-ultraviolet continuum depends on the temperature
distribution across the emitting clouds. An illustration of the results is
presented in Figure 1.
The most significant models are chosen  with 
the following criteria: shock velocities in the range 
100-300 \kms, characterizing the velocity field of the NLR, with
related pre-shock densities in the range 100-300 \cm3,
in order to reproduce the observed densities. Such \Vs ~values produce
postshock zones with temperatures of about $ \sim 10^5 - 10^6$ K.
Regarding the ionizing radiation, the characteristic
log \Fh ~generally varies between 10 and 12.7.

Assuming that starbursts can also be effective in AGN, we 
 present the results
of models with a black body temperature of 5 $10^4$ K which
are in agreement with starburst values  
and U between 0.01 and 1 (see Viegas et al. 1999).

The results are shown in Figure 1a for
a power-law radiation and in Figure 1b for the \Ts=
5 $10^4$ K blackbody
radiation, with dust-to-gas ratio equal to $10^{-15}$. 
In each figure, the left pannel shows the temperature distribution
across a cloud, where the left edge of the diagram corresponds to 
the shock front, while the right edge to the photoionized side.
The right pannel shows the corresponding SEDs.

The particular case of a blackbody spectrum with a high 
ioniziation parameter (U=100., Table 1b, left pannel)
shows that  black body radiation  
hardly heats the gas to temperatures higher
than T= $10^4$ K, since there are not  enough high energy 
photons. On the other hand, for power-law models higher
gas temperatures (T$\simeq$ 4$\times$ 10$^4$ K)
  can be reached depending on \Fh.

The peak of bremsstrahlung emission in the optical
range of  the SED depends on the temperature
distribution across the clouds. Therefore, different sets of
input parameters provides SEDs with different shapes.
The peaks,  shown in Figures 1a and 1b,
correspond to gas at $\simeq 10^4$ K, while the peaks 
in the soft X-rays are produced  by the gas at higher
temperatures in the post-shock region,
and are clearly related to the shock velocity.

\subsection{Comparison with the observations}

In Figure 2 the spectral  luminosities of all the galaxies of the sample
(open squares)
appear together. NGC 3393  is recognizable as the lower limit (crosses)
and NGC 7582 as the upper limit (open triangles). 

The references of the observational data for the continuum of the Seyfert 2 
galaxies of HS,
 from  infrared  to X-rays appear in Table 2a.
The 1.4 GHz data were taken from Heckman et al. (1995).

 A roughly common shape of the SED can be noticed.
In the radio range the data show the band of the luminosities.
Collectively, however, they cannot indicate any kind of slope, either with
bremsstrahlung or with power-law characteristics (Fermi mechanism).

On the basis of the previous analysis of the continuum and 
emission line spectra of individual Seyfert 2 galaxies,  
we have selected the least number of models which fit 
the continuum of the HS galaxies (see Table 1). The great number
of different conditions obtained by the full modeling of individual 
objects shows that modeling  a sample collectively can give only a rough idea
of the real picture.

\subsubsection{Fitting the continuum : the sample }

The observed and calculated  continuum SEDs  are compared in Figure 3.
The SED of the continuum in the optical-UV range shows a complex nature.
Notice that the data are, generally, not reddening corrected,
however, the correction is small, even in the optical-UV range
with E(B-V) between $\sim $ 0.0 and 0.1 (E(B-V)$\simeq$ 0.35 for NGC 2110, 
McAlary et al 1983).

One of the difficulties of the study of the AGN continuum is the
extraction of the stellar population component. Extraction of the 
stellar continuum from long slit spectroscopic  data of AGN are usually
obtained using a template (see, for instance, a comprehensive analysis 
by Cid-Fernandes, Storchi-Bergmann \& Schmitt 1998). For low 
dispersion data as used in this paper, however,
a less sophisticated correction is usually applied;
the stellar continuum 
is  represented by  a low temperature blackbody 
component. Here, for all the galaxies, a blackbody spectra with 
 3 $10^3$ K$\leq$ T $\leq$ 5 $10^3$ K is used to mimic the emission from the
 old-stellar galactic population which contributes to the nuclear continuum.
Model results are then used to fit data, in the same frequency range,
that are  uncontaminated by the old stellar population.
These  results correspond mainly to bremsstrahlung from gas 
photoionized by the AC radiation.
The  diffuse secondary emission from the hot slabs
of the shocked gas may also contribute to the optical-ultraviolet spectrum.

For each galaxy the models representing the main components are given in
Table 2b.
The black body temperature corresponding to the old population stars
is given in column 2 and the input parameters of the models in columns
3-7. In column 8 we give the weights (W) which are adopted to fit the data for 
each model. They represent the ratio of the emitting surface at the nebula
 to the surface at earth (4$\rm \pi \,d^2$ , where d is the distance to the 
galaxy).
 The  covering factor ($\eta$) - corresponding to the models fitting the
SED between  $10^{14}$-5 $10^{14}$ Hz - is
 calculated assuming  that  the NLR is located
 $\sim$ 1 Kpc from the central source and is listed 
 in column 9. Notice that  the models which
fit the SED at $10^{14}$-5 $10^{14}$ Hz  generally
show the highest $\eta$, thus the values listed in Table 2b 
are upper limits.

Regarding the galaxies in the sample,  Figure 3 shows that:

1) For most of the galaxies, the data available in the radio range 
are not sufficient  to indicate a definitive slope characterizing
  the emission mechanism.  However, for  NGC 2992, Mrk 3, and 
Mrk 463 the radio emission correponds to  synchrotron radiation 
produced by the Fermi mechanism  at the
shock front (Contini, Prieto, \& Viegas 1998b). On the other hand, for NGC 2110,
NGC 3393, NGC 4388, NGC 5135, NGC 5643, NGC 6221, NGC 7582, and Mrk 348 the
radio emission 
is dominated by free-free radiation that can often be reduced by 
absorption.

2) Emission in the IR is due to reradiation by dust
 (see Kraemer \& Harrington 1986).
Generally, the observational 
 data  are better explained  by a multi-cloud model  (e.g. NGC 3393, NGC 
4388).
In these cases, the dust temperature is different in the different clouds
and each component peaks at a different frequency. Thus,
the resulting peak is flatter, better reproducing the observed data.
 For
NGC 2110, NGC 3393, NGC 5506, Mrk 3, Mrk 34, Mrk 78, Mrk 348,  
IC 3639, and IC 5135 the  dust-to-gas ratio in  some of the clouds is 
particularly high  ($\sim 10^{-13}$).

3) Except for Mrk 477 and Mrk 573,  an old stellar population, 
with temperatures in the range 3 to 5 $10^3$ K, seems to be contributing
to the optical continuum of the galaxies in the sample, representing
the underlying stellar population. 

4) Soft X-ray data are fitted by gas with a  high shock velocity
( $>$ 900 \kms). This soft-X ray component originates from
the post-shock zone where \Vs $\sim$ 900 \kms, corresponding to
temperatures of $\sim$ 1.2 $10^7$ K. Such temperatures are in good agreement
with the Raymond-Smith interpretation of the ASCA  soft X-rays data
for NGC 5506 (Wang et al. 1999) and NGC 7582 (Xue et al. 1998) which
gives T $\sim 10^7$ K. Also for NGC 2110  temperatures of 6.8 $10^6$-
1.5 $10^7$ K are deduced from ROSAT soft X-ray data (Weaver et al 1995).

Dust grains  can also be heated to relatively 
high temperatures (Viegas \& Contini 1994) in the post shock region. Consequently,
besides the soft X-ray emission, the shape of the mid-IR continuum may 
provide another  key to the  presence of  high velocity clouds in AGN. In this
case we expect the mid-infrared emission to be correlated 
with the soft X-ray component. For the galaxies analysed here, 
high velocity clouds  (\Vs = 900-1000 \kms) invoked to explain 
the data in the soft X-ray and in the near IR generally have $normal$
dust-to-gas ratios ($10^{-13}-10^{-15}$).

5) The d/g ranges from 5 $10^{-16}$, in poor dusty cases, 
to $>10^{-13}$, in dusty clouds (Table 1).
Very different conditions 
can be found in different clouds of the same galaxy (e.g Mrk 477) . 
In fact, dust, which is generally
present in star forming regions, can be destroyed by sputtering and evaporation.
Therefore, the effect of the shock is crucial because sputtering and evaporation
depend on the shock velocity and on the grain temperature, respectively.

Finally, Fig. 3 shows that shock velocities of $\sim$ 200 \kms ~and preshock 
densities of $\sim$ 200 \cm3
strongly prevail in the fit of the data in the optical range.
On the other hand, the data in the optical-UV range, between
4250 \AA ~and 1200 \AA ~(corresponding to log $\nu$ = 14.85 - 15.4),
are better fitted by model 1, characterized by \Vs =100 \kms
~and a relatively high \Fh.

\subsubsection{Fitting the continuum : particular models}

\centerline{$\it Starbursts \, in \, the \, circumnuclear \, region$}

The fit of NGC 5506 UV data with  model 1
is not exact. Moreover, we have referred to the data in the
UV for all galaxies considering that the flux drops 
at log $\nu$ $>$ 15.4. A better fit to NGC 5506 can be
obtained either by a power-law (pl) dominated model with \Vs =200 \kms, \n0=50 
\cm3,
log \Fh = 9.3, D= 6 $10^{17}$ cm, and d/g= 5 $10^{-15}$
or with a black body (bb) dominated model with  \Vs= 200 \kms, \n0= 200 \cm3,
U = 0.01, \Ts = 5 $10^4$ K, D= 5 $10^{16}$ cm, and d/g = $10^{-14}$.
In Fig. 3d, the contributions of these models to the continuum 
are represented  by the long dash lines (thick) 
and the dash-dot lines (thick), 
respectively. The sum of the models will give
an even better fit to the data.
The data in the X-ray range are fitted by a model
with \Vs = 900 \kms, \n0 = 1000 \cm3, D = 8 $10^{17}$ cm,
d/g =  5 $10^{-13}$.

The bb model which fits the high frequency data represents
the case where the bb flux from the stars reaches the very
shock front of the clouds moving outwards from the AC.
In other words, the dominant starbursts  are located 
in the circumnuclear region.

\centerline{$\it Comparison \, of \, black-body \, and \, power-law \, 
dominated \, models$}

 To distinguish  starbursts from  AGN we have run
two models with shock parameters as model 5 and model 1, but with black
body radiation   with \Ts = 5 $10^4$ K  (U=1), corresponding
to a starburst and
\Ts = 2 $10^5$ K (U=10), corresponding to a "warmer" (Terlevich \& Melnick 1985), 
respectively. The results  are presented for NGC 3081 (Figure 3b).
For models with \Vs=200 \kms ~the shock prevails and there is no
great difference between the pl model and the bb model.
Diffuse radiation from the hot slabs of the gas downstream
maintains the temperature of the gas at $\sim$ 1-2 $10^4$ K.
On the other hand, in case of a low velocity shock (\Vs=100 \kms),
even a strong bb radiation flux  corresponding to a high 
temperature  cannot heat the gas enough to shift the
peak in the optical-UV to log ($\nu$) $>$ 15 (see \S 2.1).

\centerline{$\it The \, contribution \, of \, an \, intermediate \, stellar
 \, population$}

It is suggested by Heckman et al (1995) that the optical-UV SED
of the continuum can be fitted by the black body radiation 
from relatively high temperature stars. In fact,
a young stellar population is observed in some galaxies, e.g. Mrk 477
(see Heckman et al. 1997). In Figure 4  we present the fit of the
Mrk 477 continuum by  black body fluxes corresponding to different temperatures.
The three bumps correspond to  \Ts = 4,000 K (dash-dot line), 10,000 K 
(long-dash line), and 20,000 K (short-dash line). 
The second and the third one
correspond to intermediate population stars (B). 
The ratios of the weights adopted
to  fit the data are bb(4,000) : bb(10,000) : bb(20,000) = 3.2 $10^{-11}$ :
5 $10^{-23}$ : 1.6 $10^{-24}$. These are not able  to explain the observational
 evidence of starburst activity (T. Contini 1999, private communication). 
Moreover, we assume that the UV emission by a younger (T$>5 \, 10^4$ K) 
star population is absorbed by the clouds and reemitted as bremsstrahlung 
(see  for example Figure 3b).
So, we conclude that although intermediate population stars contribute to
the continuum, the  bremsstrahlung from illuminated clouds prevails.
This is also consistent with the results of line spectra calculations.

\subsection{The spectral slopes}

One of the important observational features of optical-UV spectra in Seyfert 2
galaxies is the continuum slopes (Heckman et al. 1995). 
These suggest that the featureless continuum
comes from a reddened starburst in the ranges 
1200 - 2600 \AA ~($F_{\lambda}$ $ \propto$ $\lambda^{\beta}$)
and  1910 - 4250 \AA ~($F_{\lambda}$ $ \propto$ $\lambda^{\gamma}$).
 The frequencies corresponding to 
these critical wavelengths are indicated in Figure 3 by vertical lines.

In Table 3 the spectral slopes $\gamma$  and $\beta$ obtained
from the models fitting each galaxy in the sample 
are compared with the values given by  Heckman et al. (1995, Table 1).

The agreeement is quite good, since our results depend on the fitting of the
whole continuum spectra. It can be improved
if more observational data become available in the different
wavelength ranges. 
In particular,  for NGC 2110 and MRK 34, a better agreement 
will probably be reached when 
data in the UV become available.
Interestingly, the data in the different wavelengths
correspond to different models.
Table 3 shows that low $\gamma$ values are provided by
data which correspond to a model with relatively low \Vs, low
\n0 and log \Fh = 12.7 (models 1 and 2).

One important point regarding the slopes is the position of the optical-UV
peak.  This is highly dependent on the temperature across the cloud
as shown in \S 2.1. The model dependence of  the frequency corresponding
to the peak position is
illustrated in Figure 5, where the results for various models are
plotted, with the curves shifted vertically for sake of clarity.

 Another argument favoring the origin of the ultraviolet continuum
of Seyfert 2 galaxies in  a reddened population of hot stars 
is the correlation
between the ratio of the infrared to the ultraviolet flux
($\rm L_{ir}$/\rm $\rm L_{uv}$) and 
the slope of the ultraviolet continuum (Heckman et al. 1995). These follow
the behavior of observed starbursts (Meurer et al. 1995). In Figure 6
 we show the plot for the Seyfert 2 galaxies in the sample.
For each galaxy, we plot the far-IR/UV ratio (see Table 1 of
Heckman et al.) versus the two $\beta$ values given in Table 3.
The corresponding points are close, so the correlation
is also present even if the continuum is not due to a reddened
starburst but to the NLR emission. 

Notice that dust emission is strongly coupled to gas emission
(see, for instance, Viegas \& Contini 1994) through shock and
radiation effects and that  $\rm L_{ir}$
also depends  strongly on the dust-to-gas ratio.
For both starburst and Seyfert galaxies 
the correlation shown in Figure 6 is usually used to show that the absorbed
ultraviolet radiation is reemitted in the infrared. 
For Seyfert 2 galaxies, particularly,  the relation between the central radiation source
and infrared emission may not be direct, since both shocks and
photoionization are powering the NLR, and, in our models, the observed
UV continuum is not coming from the central source. 

The results above indicate that the continuum emission 
from the NLR clouds may be another explanation
for the featureless continuum of Seyfert 2 galaxies. 
Since this component is extended, galaxies where  this component 
is dominant should show   little or no dilution of  stellar 
absorption lines as discussed by Cid-Fernandes et al. (1998). 

\section{Constraining the models : the line spectra}

In  previous sections it was found that the SED of the continuum
in the optical-UV range
is mainly reprocessed radiation from heated gas clouds. Moreover,
it was found that
reradiation from clouds photoionized by  black body 
radiation from young stars  can hardly be distinguished from that from
clouds photoionized by a power-law  radiation flux. 
Three main cases are considered:
1) power-law radiation from the AC, 2) blackbody radiation from stars with \Ts=5 
$10^4$ K
which represent the starburst case, and 3) blackbody radiation from stars with 
\Ts=1-2 $10^5$ K which represents the Terlevich \& Melnick (1985) case.
So, the interpretation
of the line spectra is  essential  for disentangling the domain of each
mechanism.

Our suggestion that the featureless continuum 
in Seyfert 2 galaxies
could be due to the NLR continuum emission can be tested by the
observed emission lines. Our previous analysis of the  Circinus galaxy
and NGC 5252 showed that a self-consistent model can only
be obtained by simultaneous fitting of the continuum and
emission-line spectra (Contini et al 1998a, 1998b). Although a 
full discussion of the emission-line spectra 
is out of the scope of this paper, it is important to show that the models
adopted to fit the continuum of the galaxies in the sample
are also consistent with the observed line ratios.

Emission-line data  for  various objects were collected from the literature.
Those for NGC 2110, NGC 2992, and NGC 5506 come from
Shuder (1980); NGC 3081 comes from Durret \& Bergeron (1986);
NGC 3393 from Diaz, Prieto, \& Wamsteker (1988); NGC 4388
from Pogge (1988); Mrk 3, Mrk 34, Mrk 78, Mrk 348, and Mrk 573
from Koski (1978); NGC 5135 and IC 5135 from Vaceli et al. (1997).

The emission-line intensity, relative to \Hb, for the  most 
indicative lines in the optical range  is listed in Table 4 for
models 2, 3, 6, 9, 12 (Table 1). Models 2(SD) and 6(SD) correspond to
models 2 and 6, respectively, but are calculated in the shock dominated (SD)
case, i.e. adopting \Fh = 0.
The minimum  and the  maximum observed values  for the galaxies referred to
above are given in the second column, whereas model results appear 
in columns 3 to 9. Data for NGC 2110, which shows line ratios
rather different than those of the other galaxies, are also
included.

Various clouds of the NLR at different physical conditions 
must contribute to the continuum, as well as to the emission-line
ratios. Therefore, multi-cloud model results obtained by averaged 
sums are also given in Table 4 (columns 10-12). 
 The weights adopted
in the averaged sums appear in the bottom of Table 4, as well as the
calculated absolute values of \Hb ~for the individual models.
Notice that in order to have a good fit, 
the results of SD models  are included.
The weights of SD models are higher than those of radiation
dominated models because the absolute fluxes are weaker
(see Viegas et al. 1999).
The averaged results are within the maximum and minimum observed
ratios except for the [N I] emission-line, which is highly
dependent on the geometrical depth. Concerning the [N II] 6584+ line,
a slighter higher N/H abundance could provide better agreement
(cf. Contini et al. 1999).

For sake of consistency, we present in Figure 7 the SEDs corresponding
to the theoretical models AV1, AV2, and AV3.
An  hypothetical bb emission from the background old
star population is also shown (long dash-dot lines)
to better understand the diagrams.
The SED maxima for shock-dominated models 
are determined by the shock velocity, whereas
the optical-UV peaks in radiation-dominated models depend on
the radiation flux.The model results are compared to the
NGC 5643 data which are representative of the continuum
shown by the galaxies in the sample.
Notice that AV2 gives a better fit than AV1 and AV3, both
in the near infrared and in the far ultraviolet.
This suggests that the fit of the line spectra must include
shock dominated models.

\section{Concluding Remarks}

The aim of this paper is to understand
the continuum SED of Seyfert galaxies.
We show that composite models for the NLR
of Seyfert 2 galaxies can explain the full range
of the observed continuum, and, in particular, the
optical-ultraviolet continuum. 
Comparison of theoretical results and 
observational data shows that \Vs ~of
about 100-300 \kms, and \n0 of 200-300 \cm3  must prevail
in the NLR.
Higher velocities may also be  present in order to explain 
the soft X-ray emission.

Multi-cloud models are necessary to interpret the spectra 
(both line and continuum) of single objects, 
even in the global investigation of a sample of galaxies.

An important point is  the characteristics of the featureless continuum
of Seyfert 2 galaxies. Regarding the continuum slopes and the
correlation between the  far-infrared to ultraviolet ratio
and the UV slope, both  are reproduced by our models.
This is an indication that the NLR continuum emission may be the
main components of featureless continuum.

The main result of our investigation is that the
continuum observed in the HS sample is reprocessed radiation
from the clouds of the NLR. These clouds are mainly powered
by the central radiation, usually characterized by a power-law
ionizing spectrum. Nevertheless, black body radiation from starbursts
located in the outskirts of the nuclear region may, in some cases,
contribute to the UV data.
The results will be confirmed when further data in the far-UV
become available.

\bigskip

\noindent
{\bf Acknowledgements}. 
We are grateful to the referee for enlightening comments
and to G. Drukier for reading the manuscript.
This paper is partially supported
by the Brazilian financial agencies: FAPESP (1997/13816-4), CNPq
(304077/77-1), and PRONEX/Finep(41.96.0908.00).

\newpage

{\bf References}

\bigskip

\vsize=26 true cm
\hsize=16 true cm
\baselineskip=18 pt
%
\def\ref {\par \noindent \parshape=6 0cm 12.5cm 
0.5cm 12.5cm 0.5cm 12.5cm 0.5cm 12.5cm 0.5cm 12.5cm 0.5cm 12.5cm}

\ref Aaronson,M. et al. 1981, MNRAS, 195, 1;

\ref Allen, C.W. 1973 in "Astrophysical Quantities" (Athlon) 

\ref Allen,D.A. 1976, ApJ, 207, 367;

\ref Becker,R.M., White,R.L., \& Edwards, A.L. 1991, ApJS, 75, 1;

\ref Boroson,T.A.,Strom,K.M., \& Strom,S.E. 1983, ApJ, 274, 39;

\ref Cid-Fernandee, R., Dottori, H., Gruenwald, R. \& Viegas, S. M.
 1991, MNRAS 255, 165

\ref Cid-Fernandes, R., Storchi-Bergmann, T. \& Schmitt,H. 1998,
MNRAS, 297, 579

\ref Contini,M., Prieto,M.A., \& Viegas,S.M. 1998a, ApJ, 492, 511

\ref Contini,M., Prieto,M.A., \& Viegas,S.M. 1998b, ApJ, 505, 621

\ref Contini,M., Radovich,M., Rafanelli,P., \& Richter,G. 1999, submitted

\ref De Vaucouleurs, A. \& Longo, G. 1988, Catalogue of Visual and Infrared 
Photometry
of Galaxies from 0.5 $\mu$m to 10 $\mu$m (1961-1985);

\ref De Vaucouleurs, G. et al. 1991 Third Reference Catalogue of Bright 
Galaxies,

\ref Diaz,A.I., Prieto, M.A., \& Wamsteker,W.  1988, A\&A, 195, 53  

\ref Doroshenko,V.T. \& Terebezh,V.Yu 1979, SvAL, 5, 305;

\ref Drain,B.T., \& Lee,H.M. 1994, ApJ, 285, 89

\ref Durret,F. \& Bergeron,J. 1986, A\&A, 156, 51

\ref Fabbiano,G., Kim, D.-W.,\& Trinchieri, G. 1992, ApJS, 80, 531;

\ref Frogel,J.F., Elias,J.H., \& Phillips,M.M. 1982, ApJ, 260, 70;

\ref Glass,I.S. 1973, MNRAS, 164, 155;
\ref Glass,I.S. 1976, MNRAS, 175, 191;
\ref Glass,I.S. 1978, MNRAS, 183, 85;
\ref Glass,I.S. 1979, MNRAS, 186, 29;
\ref Glass,I.S. 1981, MNRAS, 197, 1067;
\ref Glass,I.S. et al. 1982, A\&A, 107, 276;
\ref Gower,J.F.R., Scott,,P.F., \& Wills,D. 1967, MmRAS, 71, 49;

\ref Gregory,P.C. \& Condon,J.J. 1991, ApJS, 75, 1011;

\ref Gregory,P.C. et al. 1994, ApJS, 90, 173;

\ref Griersmith,D., Hyland, A.R., \& Jones, T.J. 1982, AJ, 87, 1106;

\ref Griffith,M.R. et al. 1994, ApJS, 90, 179;

\ref Griffith, M.R. 1995, ApJS, 97, 347;

\ref Heckman,T., Krolik,J., Meurer,G., Calzetti,D., Kinney,A.,
Koratkar,A., Leitherer,C., Robert,C., \& Wilson,A. 1995, ApJ, 452, 549

\ref Heckman,T.M. et al. 1997, ApJ, 482, 114

\ref Joyce,R.R. \& Simon, M. 1976, PASP, 88, 870;

\ref Kinney,A.I. et al. 1993, ApJS, 86, 5;

\ref Kormendy,J. 1977, ApJ, 214, 359;

\ref Koski,A.T.  1978, 223, 56

\ref Large,M.I. et al. 1981, MNRAS, 194, 693;

\ref Lauberts,A. \& Valentijn,E.A. 1989, The Surface Photometry Catalogue
of the ESO-Uppsala Galaxies, 1989, Garching Bei Munchen ESO;
\ref Leitherer,C., Robert,C., \& Heckman, T. 1995 ApJS, 99, 173  

\ref Maddox,S.J. et al. 1990, MNRAS, 243, 692;

\ref Mathewson,D.S. \& Ford, V.L. 1996, ApJS, 107, 97;

\ref McAlary,C.W.,McLaren,R.A., \& Crabtree,D.R. 1979, ApJ, 234, 471;

\ref McAlary,C.W. et al. 1983, ApJS, 52, 341;

\ref Meurer, G., Heckman, T., Leitherer, C., Kinney, A., Robert, C.,
\& Garnett, D. 1995, AJ, 110, 2665 

\ref Moshir, M. et al. 1990, Infrared Astronomical Satellite Catalogs, 1990,The 
Faint Source Catalog, Version 2.0;

\ref Mould,J.,Aaronson,M.,\& Huchra,J. 1988, ApJ, 238, 458;

\ref Neugebauer,G. et al. 1976, ApJ, 205, 29;

\ref Pogge,R.W.  1988, ApJ, 332, 702

\ref Rieke,G.H. 1978, ApJ, 226, 550;

\ref Rieke,G.H. \& Low,F.J. 1972, ApJ, 176L, 95;

\ref Rudy,R.J.,Levan,P.D., \& Rodriguez-Espinosa,J.M. 1982

\ref Sandage,A. \& Visvanathan,N., 1978, 223,707; 

\ref Scoville,N.Z. et al. 1983, ApJ, 271, 512;

\ref Shuder,J.M. 1980, ApJ 240, 32

\ref Soifer,B.T. et al. 1989, AJ, 98, 766;

\ref Stein,W.A. \& Weedman,D.W. 1976, 205, 44;

\ref Terlevich, R. \& Melnick, J. 1985, MNRAS, 213, 841

\ref Tran, H. D. 1995, ApJ, 440, 578

\ref Vaceli,M.S., Viegas,S.M., Gruenwald,R., \& De Souza,R.E.   1997,  AJ, 114,
1245

\ref Viegas,S.M. \& Contini,M. 1994, ApJ, 428, 113

\ref Viegas,S.M., Contini,M., \& Contini,T. 1999, A\&A, 347, 112

\ref Wang,T. et al. 1999, ApJ, 515, 567

\ref Ward,M. et al. 1982, MNRAS, 199, 953;

\ref Weaver,K.A. et al. 1995, ApJ, 442, 597

\ref White,R.L. \& Becker,R.H. 1992, ApJS, 79, 331;

\ref Wright,A.E. et al 1996, ApJS, 103, 145;

\ref Wright,A.E. et al. 1994, ApJS, 91, 111;

\ref Wright,A. \& Otrupcek,R. 1990, Parkes Catalogue, 1990, Australia Telscope 
National Facility;

\ref Xue,S.-J. et al. 1998, PASJ, 50, 519

\newpage

\hsize=14 true cm

{\bf Figure Captions}

Fig. 1

\noindent
The results  corresponding to
a power-law ionizing radiation (a) and  a 5 $10^4$ K blackbody
radiation (b). In each figure, the left pannel show the temperature distribution
across a cloud, where the left edge of the diagram corresponds to 
the shock front, while the right edge to the photoionized side. 
The  thin vertical   line in the middle of the diagrams
indicates the separation of the cloud in two halves. The axis scales are
logarithmic. The horizontal axis scale is symmetric in order to 
provide an equal view of the
two sides of the cloud, that dominated by collisional ionization and the
radiation dominated one.
The right pannel shows the corresponding spectral energy distribution.
The power-law results refer to \Vs = 100 \kms and \n0 = 100 \cm3 
(thin lines) and to \Vs = 300 \kms and \n0 = 300 \cm3 
(thick lines). Solid , short-dashed, and long-dashed lines correspond 
to log\Fh = 12, 11, and 10 , respectively.
The black body results were obtained for U=0.01
(long-dashed), 0.1 (short-dashed), and 1. (solid), and  U=100. 
(dash-dot line).

Fig. 2
 
\noindent
The spectral luminosities of all the galaxies of the
 Heckman et al. (1995) sample.

Fig. 3

\noindent
The fit of the calculated to the observed SED  for all the objects
of the Heckman et al. sample.
Solid lines : models 4, 11, 12, and 13; short dashed lines : models 5, 6, and 7 
;
long dashed lines : models 1, 2; short dash-dot lines : models 3 and 8; 
dotted lines : models 14, 15 and 16; long dash-dot : the black body emission
from the back ground old star population.  Filled squares represent the data. 
Vertical thin lines define the crucial wavelengths at
4250 \AA, 2600 \AA, 1910 \AA, and 1200 \AA  (see \S 2.3).

Fig. 4

\noindent
The fit of MRK 477 continuum SED by star population
at different temperatures.
The three bumps between 14 $<$ log($\nu$) $<$ 15
 correspond to black body radiation with  T$_*$ = 4,000 K (dash-dot line), 
10,000 K 
(long-dash line), and 20,000 K (short-dash line). 
Solid lines correspond to bremsstrahlung emission from the gas  
and to IR thermal  emission
from dust from the NLR clouds (see Figure 3i).

Fig. 5

\noindent
Comparison of the  optical-UV peaks in the SED of the continuum calculated
by models presented in Figs. 1a and 1b.
1 : \Vs=100 \kms, \n0= 100 \cm3, log\Fh=12;
2 : \Vs=100 \kms, \n0= 100 \cm3, log\Fh=11;
3 : \Vs=300 \kms, \n0= 300 \cm3, log\Fh=12;
4 : \Vs=300 \kms, \n0= 300 \cm3, U=1;
5 : \Vs=300 \kms, \n0= 300 \cm3, U=0.1;
6 : \Vs=300 \kms, \n0= 300 \cm3, U=0.01;
7 : \Vs=100 \kms, \n0= 100 \cm3, U=1;
8 : \Vs=100 \kms, \n0= 100 \cm3, U=0.1;
9 : \Vs=500 \kms, \n0= 500 \cm3, U=1;

Fig. 6

\noindent
The far-infrared to ultraviolet luminosity ratio  ($\rm L_{ir}$/$\rm L_{uv}$) 
versus the slope
of the ultraviolet continuum $\beta$ (see \S 2.3). The circles refer to our 
results and
the stars to Heckman et al. (1995), as listed in Table 3. The straight
line represents the correlation obtained from starburst data (Meurer et
al. 1995).

Fig. 7

\noindent
The SED of the continua referring to the multi-cloud models AV1, AV2, and AV3
(Tables 4). Shock dominated models 
are represented by dotted lines, model 3 by short dash-dot, model 6 by
long dash, and model 12 by solid lines. Model 2 has a low weight and does 
not appear in the figures.
An  hypothetical bb emission from the background old
star population is also shown (long dash-dot lines).
Filled squares  refer to NGC 5643 data.

\newpage

\oddsidemargin 0.1cm
\evensidemargin 0.1cm

\begin{table}
\centerline{Table 1}
\centerline{Input parameters for the models}
\begin{center}
\begin{tabular}{ l l l l l l   } \\ \hline\\
 model & \n0 & \Vs & log(\Fh$^o$) & D & d/g  \\
  (1) & (2) & (3) & (4) & (5) & (6) \\ 
\hline\\
 1       & 150 & 100 & 12.7 & 3(16) & 5(-15)  \\
 2        & 150 & 100 & 12.7 & 3(16)  & 3(-14)  \\
 3       & 400 & 100 & (1.5,5.0)* & 3(16) &5(-15) \\
 4       & 400 & 150 & 12 & 3(16) & 5(-17)  \\
 5       & 200 & 200 & 11.7 & 3(18) & 5(-16)  \\
 6       & 200 & 200 & 11 & 3(18) & 1(-15)  \\
 7       & 200 & 200 & 11 & 3(19) & 7(-16)  \\
 8       & 200 & 200 & (0.1,5.3)* & 4.5(17) & 5(-15) \\
 9       & 200 & 200 & (1.,4.7)* & 3(18)& 1(-15) \\
 10      & 200 & 200 & (0.01,4.7)* & 5(16) & 1(-14) \\    
 11      & 200 & 300 & 12 & 3(18) & 5(-15)  \\ 
 12      & 200 & 300 & 12 & 3(18) & 1(-13)  \\
 13      & 100 & 400 & 12 & 5.5(18) & 5(-13)\\
 14      &  200 & 900 & SD   & 8(17) & 5(-13)\\
 15      & 1000 & 1000 & SD  & 3(18)&2(-14) \\
 16      & 1000 & 1000 & SD  & 3(18) & 5(-15)  \\
 &&&&&\\ \hline\\

\end{tabular}

\end{center}

$^o$
U=\Fh/(nc($\alpha$-1)) ($\rm (E_H)^{- \alpha+1}$ - $\rm (E_c)^{- \alpha+1}$)

($\rm E_c$= high energy cutoff = 3000 eV, $\rm E_H$=ionization potential
of H.

* Blackbody radiation characterized by (U,log(\Ts))
\end{table}

\newpage

\begin{table}
\centerline{Table 2a}
\centerline{The references of the continuum data}
\begin{center}
\small{
\begin{tabular}{ l l   } \\ \hline\\
 galaxy& references  \\
 \hline\\
 NGC 2110 &1, 2, 3, 4, 5\\
 NGC 2992 &1, 2, 3, 6, 7, 8, 9, 10\\
 NGC 3081 &3, 6, 7, 11, 12, 13, 14\\
 NGC 3393 & 4, 7, 10, 11, 12\\
 NGC 4388 &4 ,6, 7, 11, 15, 16, 17, 18, 19, 20, 21\\
 NGC 5135 &4, 6, 7, 11, 12, 22\\
 NGC 5506&1, 2, 3, 4, 6, 7, 8, 11, 19, 24, 25\\
 NGC 5643 &4, 6, 7, 11, 12, 15, 26, 27\\
 NGC 5728 &4, 6, 7, 11, 19, 28\\
 NGC 6221  &1, 7, 9, 11, 12, 15, 26, 27\\
 NGC 7582 & 1, 2, 4, 6, 7, 8, 9, 11, 12, 15, 27, 29, 30, 31, 32, 33, 34, 35\\
 MRK 3  &4, 7, 15, 20, 21, 28, 36, 37, 38, 39\\
 MRK 34  &4, 36, 37, 40\\
 MRK 78  &4, 36 \\
 MRK 348 & 4, 7, 11, 15, 20, 21, 25, 36, 38, 45\\
 MRK 463 & 4, 7, 11, 20, 21, 41, 42\\ 
 MRK 477 & 4, 7, 11, 43, 44\\
 MRK 573 &4, 7, \\
 IC 3639 &3, 4, 7, 11, 12\\
 IC 5135 & 4, 7, 11, 12\\
 &\\ \hline\\

\end{tabular}}

\end{center}

1:McAlary et al. 1983;
2:Glass 1981;
3:Ward et al. 1982;
4:Moshir et al. 1990;
5:Griffith 1995; 6:Fabbiano, Kim, \& Trinchieri 1992;
7:De Vaucouleurs et al. 1991;
8:Glass 1979;
9:Glass et al. 1982;
10:Griffith et al. 1994;
11:Kinney et al. 1993;
12:Lauberts \& Valentijn 1989;
13:Kormendy 1977;
14:Sandage \& Visvanathan 1978;
15:De Vaucouleurs \& Longo 1988;
16:Boroson, Strom, \& Strom 1983;
17:Mould, Aaronson,\& Huchra 1988;
18:Scoville et al. 1983;
19:Soifer et al. 1989;
20:Gregory \& Condon 1991;
21:Becker, White, \& Edwards 1991;
22:Wright et al. 1996;
23:Maddox et al. 1990;
24:Glass 1978;
25:White \& Becker 1992;
26:Gregory et al. 1994;
27:Wright et al. 1994;
28:McAlary, McLaren, \& Crabtree 1979;
29:Mathewson \& Ford 1996;
30:Griersmith, Hyland,  \& Jones,  1982;
31:Glass 1973;
32:Glass 1976;
33:Aaronson et al. 1981;
34:Frogel, Elias, \& Phillips 1982;
35:Wright \& Otrupcek 1990;
36:Rieke 1978;
37:Neugebauer et al. 1976;
38:Joyce \& Simon 1976;
39:Gower, Scott, \& Wills 1967;
40:Rieke \& Low 1972;
41:Doroshenko \& Terebezh 1979;
42:Large et al. 1981;
43:Rudy, Levan \& Rodriguez-Espinosa 1982;
44:Allen 1976;
45:Stein \& Weedman 1976;

\end{table}

\begin{table}
\centerline{Table 2b}
\centerline{The best fitting models}
\begin{center}
\tiny{
\begin{tabular}{ l l l l l l lll  } \\ \hline\\
 galaxy&  $\rm T_{old}$& \n0 & \Vs & log(\Fh) & D & d/g & log(W) &$\eta$  \\
  (1) & (2) & (3) & (4) & (5) & (6) & (7)&(8)&(9) \\ 
 \hline\\
 NGC 2110 &3000&-&-&-&-&-& -18.9&-\\
         && 150 & 100 & 12.7 & 3(16)  & 3(-14)&-12.1 &-  \\
         & & 200 & 200 & 11.7 & 3(18) & 5(-16)&-10.35&0.088   \\
         & &200 & 300 & 12 & 3(18) & 1(-13) &-11.3&-   \\
 NGC 2992 &4000&-&-&-&-&-&-19.1&-\\
         & & 200& 200 & 11.7&3(18) & 5(-16)&-10.5&0.07   \\ 
          & &1000 & 1000 & s.d.& 3(18)&2(-14)&-12.9&-      \\
 NGC 3081 &4000&-&-&-&-&-&-19.3&-\\
         & & 150 & 100 & 12.7 & 3(16) & 3(-14)&-11.2 &-     \\
          && 150 & 100 & (10,5.3) & 3(16) & 3(-14)& -9.2&- \\
         & &200 & 200 & 11.7&3(18) & 5(-16)&-10.6&0.06   \\
         &&200 & 200 & (1.,4.7)&3(18)&5(-16)&-10.6&- \\
         & &1000 & 1000 & s.d.& 3(18)&2(-14)&-14.6&-      \\
 NGC 3393 & 4000&-&-&-&-&-& -20.35&-\\
         & &200 & 200 & (0.1,5.3) & 4.5(17) & 5(-15)&-11.1&- \\
         && 200 & 200 & 11 & 3(19) & 7(-16)&-10.7&0.115  \\
         && 100 & 400 & 12 & 5.5(18) & 5(-13)&-12.7&- \\
         &&  200 & 900 & s.d. & 8(17) & 5(-13)&-12.4&- \\
 NGC 4388 &4000& -&-&-&-&-&-18.5&-\\
         & &150 & 100 & 12.7 & 3(16) & 5(-15)& -10.95 &-   \\
         & &200 & 200 & 11.7&3(18) & 5(-16) &-9.7&0.52  \\
         & &1000 & 1000 & s.d.& 3(18)&2(-14) &-14.4 &-   \\
 NGC 5135 &4000 &-&-&-&-&-&-19.&-\\
         && 150 & 100 & 12.7 & 3(16) & 3(-14)& -10.75&-   \\
         & &200 & 200 & 11 & 3(18) & 1(-15) &-9.1&1.    \\
         & &1000 & 1000 & s.d.& 3(18)&2(-14)&-14.7&-   \\
 NGC 5506&3000&-&-&-&-&-&-18.4&-\\
         && 200 &  50 & 9.3  & 6(17) & 5(-15)&-8.&-   \\
         & &200 & 200 & 11 & 3(18) & 1(-15)  &-10.&0.14   \\
         & &200 & 200 & (0.01,4.7) & 5(16)&1(-14)&-8.&- \\
         & &200 & 300 & 12 & 3(18) & 1(-13)& -10.8&-     \\
         && 1000 & 1000 & s.d.& 3(18)&2(-14)&-12.8&-   \\
 NGC 5643 &4000 &-&-&-&-&-&-18.5&-\\
          &&150 & 100 & 12.7 & 3(16) & 5(-15)&-11.3 &-    \\
          && 150 & 100 & 12.7 & 3(16) & 3(-14)&-11.4&0.003      \\
        & &200 & 200 & 11 & 3(18) & 1(-15)&-9.3& 0.3   \\
         &&1000 & 1000 & s.d. & 3(18) & 2(-14)&-13.3&-   \\
 NGC 5728 &5000&-&-&-&-&-&-19.5&-\\
         && 150 & 100 & 12.7 & 3(16) & 3(-14)&-11.2&-   \\
         & &200 & 200 & 11 & 3(18) & 1(-15)&-10.4&0.13     \\
         && 1000 & 1000 & s.d.& 3(18)&2(-14) &-14.6&-    \\
 NGC 6221  &5000&-&-&-&-&-&-19.0 &-\\
         &&150 & 100 & 12.7 & 3(16) & 5(-15)&-11.4 &-  \\
          & &200 & 200 & 11.7&3(18) & 5(-16)&-10.0 &0.09 \\
 NGC 7582 & 5000&-&-&-&-&-&-19.2&-\\
         &&150 & 100 & 12.7 & 3(16) & 3(-14)&-11.2 &-     \\
          & &200 & 200 & 11.7&3(18) & 5(-16)&-9.&1.   \\
         & &1000 & 1000 & s.d.& 3(18) & 5(-15)&-13.2&-  \\
 MRK 3  &4000&-&-&-&-&-& -19.6&-\\
        &&200 & 300 &12&3(18) & 1(-13)&-10.8&0.11\\
 MRK 34  &4000&-&-&-&-&-&-20.2&-\\
         &&200 & 300 & 12 & 3(18) & 1(-13)&-11.5 &0.3  \\
 MRK 78  &4000&-&-&-&-&-&-20.4&-\\
         &&200 & 300 & 12 & 3(18) & 1(-13)&-11.5 &0.16 \\
 MRK 348 & 4000&-&-&-&-&-&-19.7&-\\
         &&400 & 100 & (1.5,5.0) & 3(16) &2(-16) &-9.5&-    \\
         && 200 & 200 & 11.7&3(18) & 5(-16)&-10.8 &0.13  \\
         & &200 & 300 & 12 & 3(18) & 1(-13)&-11.2 &0.05   \\
 MRK 463 & 4000&-&-&-&-&-&-19.7&-\\ 
         &&200 & 300 & 12 & 3(18) & 1(-13)&-10.95&0.33     \\
          && 150 & 100 & 12.7 & 3(16)  & 3(-14)&-11. &-    \\
         & &1000 & 1000 & s.d.& 3(18)&2(-14)&-12.7&-   \\
 MRK 477 & -&-&-&-&-&-&-&-\\
         &&150 & 100 & 12.7 & 3(16) & 3(-14)&-8.9& 1. \\
         & &200 & 300 & 12 & 3(18) & 1(-13) &-11.45&0.19    \\
 MRK 573 &-&-&-&-&-&-&-&-\\
        & &400 & 100 & (1.5,5.0) & 3(16) &2(-16)&-8.9     \\
         & &200 & 200 & 11.7&3(18) & 5(-16)&-10.6 &0.26  \\
 IC 3639 & 4000&-&-&-&-&-&-19.2&-\\
         &&150 & 100 & 12.7 & 3(16) & 5(-15)& -10.8&- \\
         & &200 & 300 & 12 & 3(18) & 1(-13)&-11. &0.04    \\
 IC 5135 & 4000-&-&-&-&-&-&-19.2&-\\
         &&150 & 100 & 12.7 & 3(16) & 5(-15)&-10.9&-   \\
          & &150 & 100 & 12.7 & 3(16)  & 3(-14)&-11.4&-     \\
         & &200 & 300 & 12 & 3(18) & 1(-13)&-10.8&0.12     \\
 &&&&&&\\ \hline\\

\end{tabular}}

\end{center}
\end{table}

\begin{table}
\centerline{Table 3}
\centerline{The $\gamma$ and $\beta$ indices}
\begin{tabular}{l llll l ll l}\\ \hline   \\
 galaxy&$\gamma$ (HS)&  $\gamma$ (mod)& $\beta$ (HS) & $\beta$ (mod)\\
  & &&&  \\
  (1) & (2) & (3) & (4) & (5)  \\ 
\hline\\
\ NGC 2110 &0.7& 1.3*&-&- \\
\ NGC 2992 &1.0&1.16&-&- \\
\ NGC 3081 &1.2& 0.4 & -0.4 &-0.5 \\
\ NGC 3393 &1.5 &1.88&-0.4&-0.7 \\
\ NGC 4388 &1.0&1.0&-0.9&-0.6  \\
\ NGC 5135 &0.4& 0.5& -0.1 &-0.1 \\
\ NGC 5506 &0.9&1.16 &-0.4&-0.7 \\
\ NGC 5643 &2.&1.6 &-0.3&-0.5 \\
\ NGC 5728 & 1.4 &1.0 &-0.6&-0.4 \\
\ NGC 6221& 1.0&1.59&0.4& 0.49 \\
\ NGC 7582 &2.5&2.5&1.7&1.38 \\
\ MRK 3 &1.6&1.88&-0.3&- \\
\ MRK 34 &0.6&1.59*&-0.6&-  \\
\ MRK 78 &1.4&1.8 &0.0&- \\
\ MRK 348 &0.2&0.58 &0.1&0.19 \\
\ MRK 463 &-0.2&-0.42&-0.7&-0.4 \\
\ MRK 477 &-0.6&-0.3&-1.3&-1.0 \\
\ MRK 573 &0.9&1.3 &-0.7 &- \\
\ IC 3639 &0.3&0.2&-1.0 &-1.7 \\
\ IC 5135 &0.9&0.43 &-0.5&-0.43 \\  
\hline\\
\end{tabular}

* UV data are not available
\end{table}

\begin{table}
\centerline{Table 4}
\centerline{The average emission-line spectra}
\begin{center}
\small{
\begin{tabular}{ l l l l l l  llllll } \\ \hline\\
 emission-line  & obs    &2(SD)& 6(SD)&2 & 3& 6 & 9 & 12 &AV1   &AV2    &AV3     
  \\
  (1) & (2) & (3) & (4) & (5) & (6) & (7) & (8) & (9) & (10) & (11) & (12) \\ 
\hline\\
\ [OII] 3727 & 1.9-6.4&72. & 40.& 0.& 5(-4)&1.2 & 1.6 & 0.04&6.35&2.0 &3.36\\
\ [NeIII] 3869 & 0.7-2.3&3.0 & 3.3 & 4(-4)& 0.3&0.66&  0.68& 1.0 
&0.92&0.7&0.98\\
\ [OIII] 4363 & 0.13-0.35&2.25& 2.0 & 1(-4) & 0.03 &0.05& 0.04 & 0.1 &0.29&0.1 
&0.21\\
\ HeII 4686 & 0.06-0.48&0.02 & 0.1 & 1.0 & 1.1 &0.086& 0.027 & 0.5 
&0.41&0.44&0.49\\
\ [OIII] 5007+&6.4-16.0&28.2 & 24.8 & 0.01 & 1.9 & 10. & 9.8 & 18.&11.1& 9.7 
&15.\\
\ [NI] 5200 & 0.07-0.24 & 0.16 & 0.09 & 0.0 & 0.0 &0.020& 0.028 & 
0.0&0.02&0.013&0.01\\
\ HeI 5876 & 0.05-0.19 &0.15 & 0.2 &0.0&0.001 &0.13 & 0.19 & 0.08 
&0.1&0.085&0.08\\
\ [FeVII] 6086 & 0.01-0.19&0.003 & 0.003 & 0.0 & 0.61 & 5(-4)& 1(-3) & 
1(-5)&0.15& 0.17 & 0.09\\
\ [OI] 6300+ & 0.3-1.6&0.46 & 0.5 & 0.0 & 0.0 &0.87 & 1.12 & 1(-5)&0.46&0.46& 
0.18\\
\ [NII] 6584+& 2.0-6.1&8.6 & 6.8 & 0.0 & 3(-4) &2.50 & 3.90 & 0.018 
&2.0&1.5&0.9\\
\ [SII] 6717 &0.7-1.6 &2.3 & 1.28 & 0.0 & 0.0 &1.24& 1.54 & 0.6& 0.86&0.8 
&0.68\\
\ [SII] 6731 & 0.1-1.6 & 2.5 & 1.96 & 0.0 & 0.0 &1.78& 2.00 & 1.2 & 1.27&1.2 
&1.17\\
\ \Hb ~(\erg) &-& 7.(-5)&2.2(-4) & 1.8(-3) & 0.67 &0.75& 0.38 & 13.5 &-&-&- \\
\ W(AV1)&-& 4000 & 3000 & 0.1 & 3.0 &5.0&- & 0.1 &-&-&-\\
\ W(AV2)&-& 2000& 10 & 0.1 & 3.0 &5.0&-& 0.1 &-&-&-\\
\ W(AV3)&-& 1000 & 1000 & 0.1 & 1.0 & 1.0 &-& 0.2&-&-&- \\
\hline \\
\end{tabular}}
\end{center}
\end{table} 

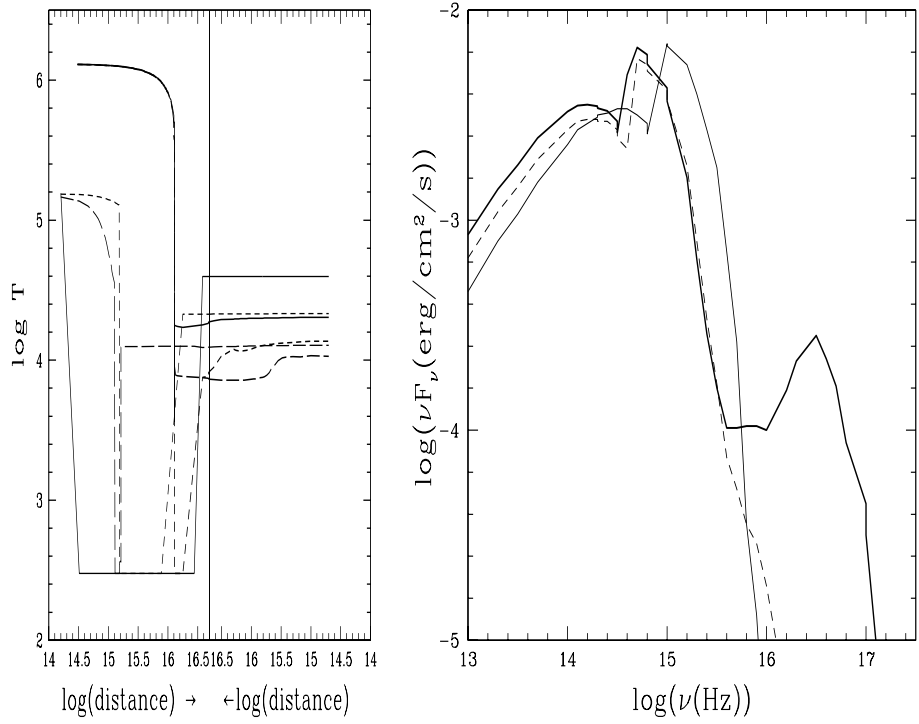
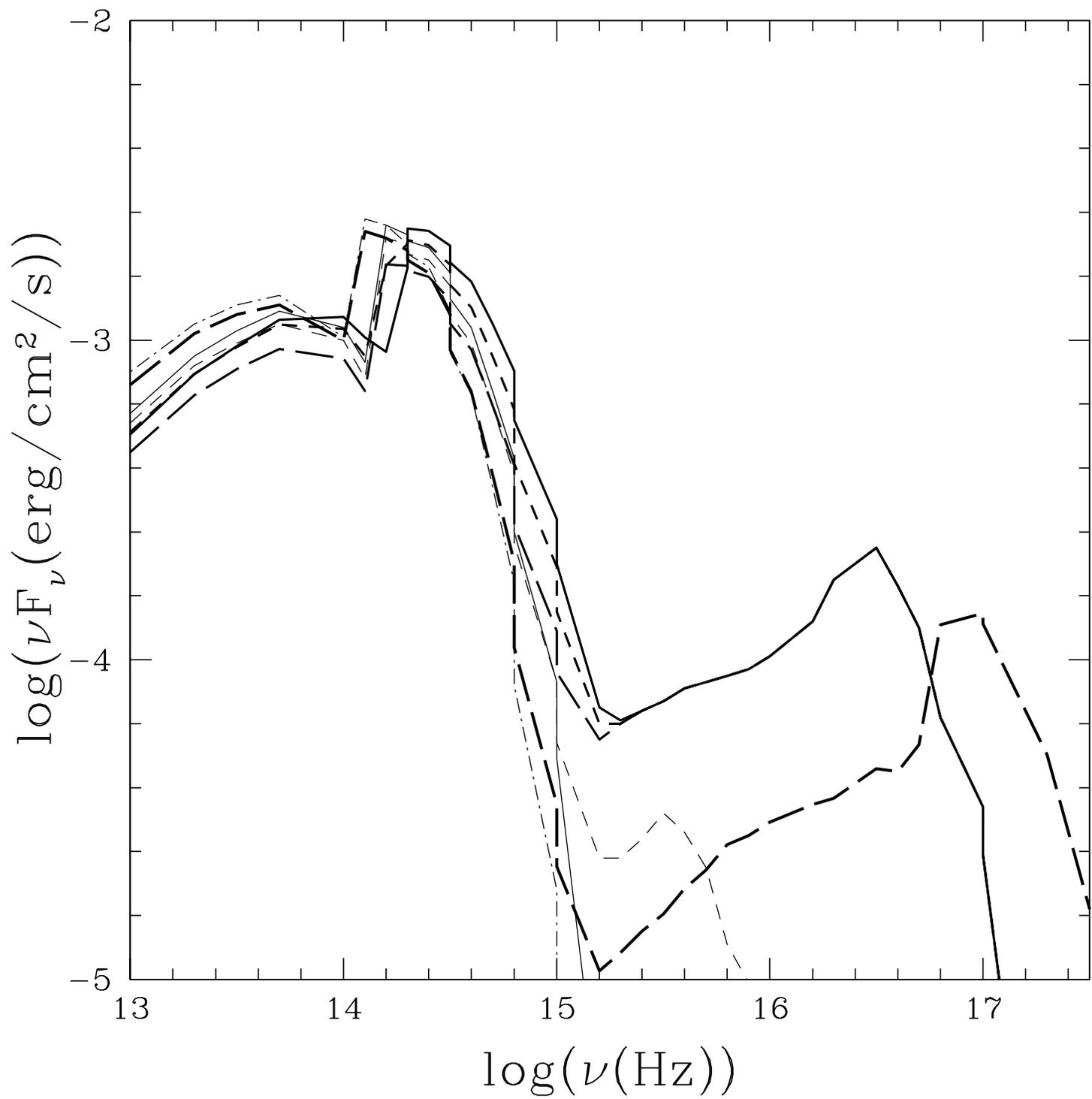
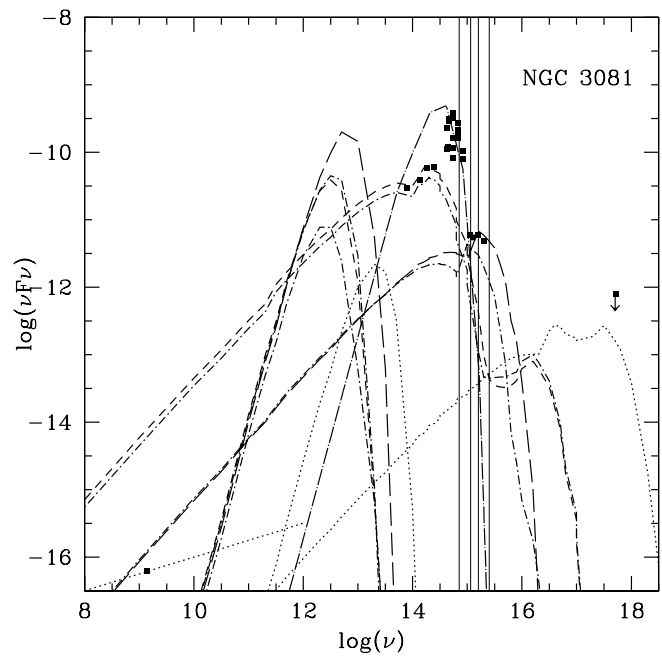
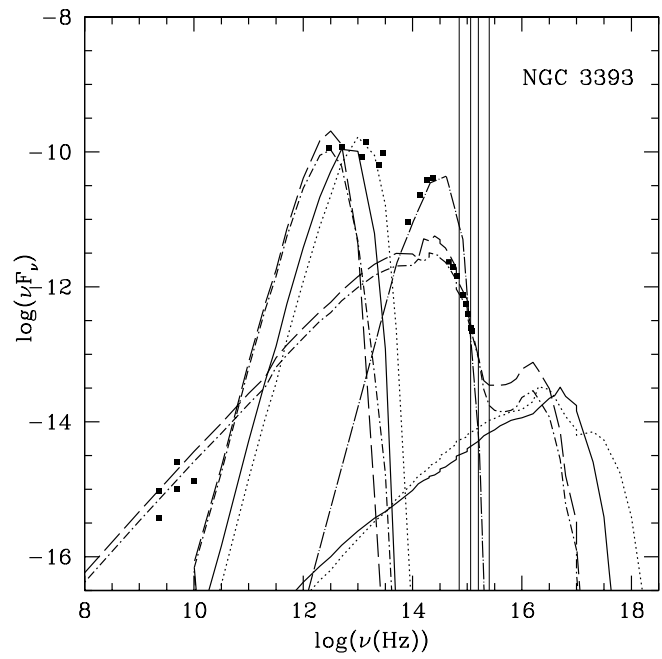
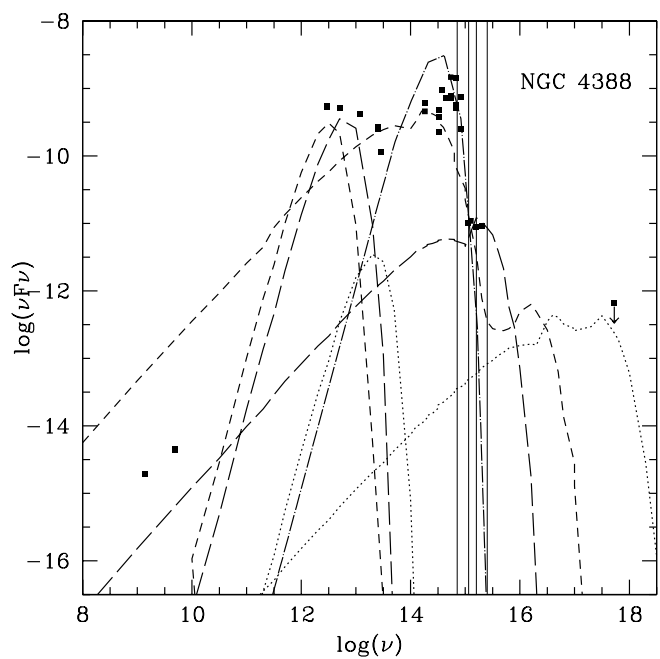
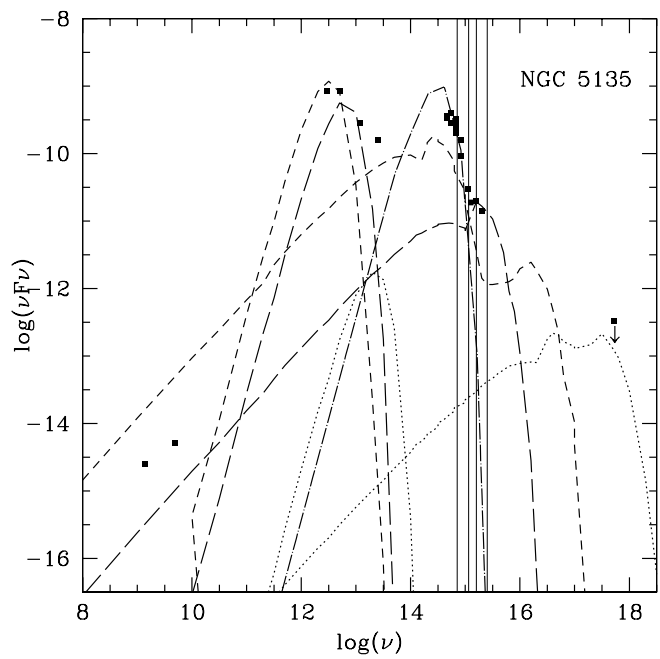
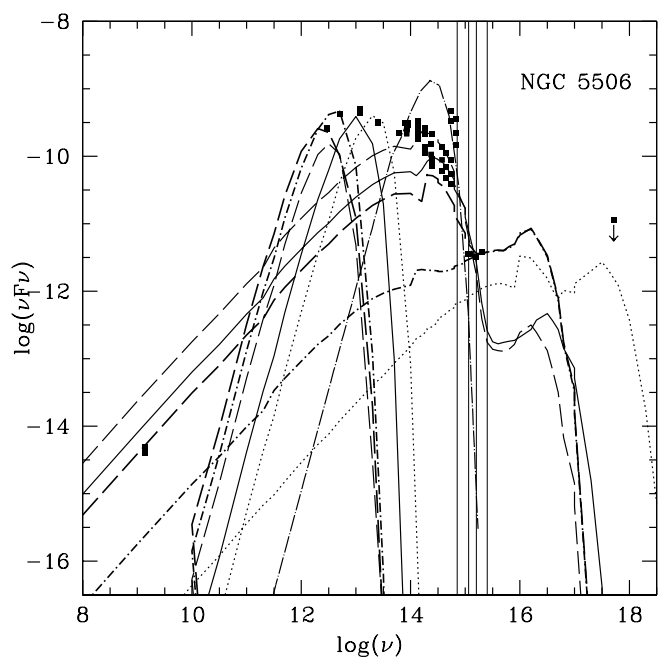
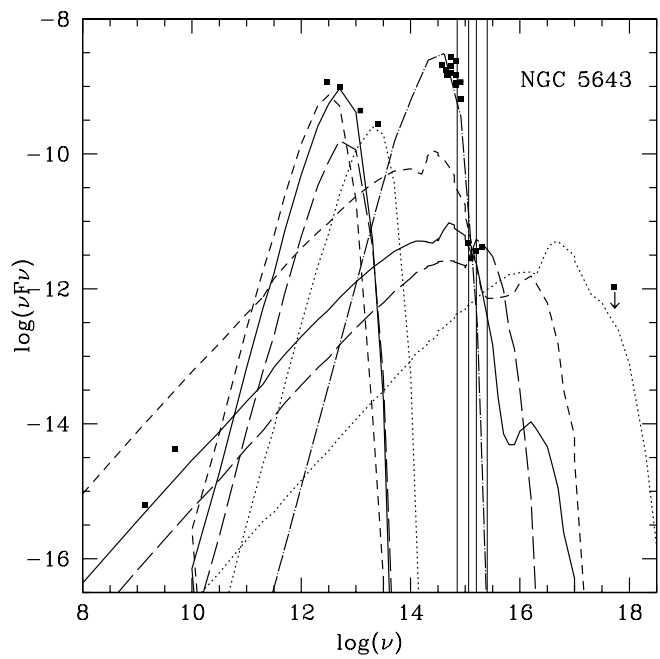
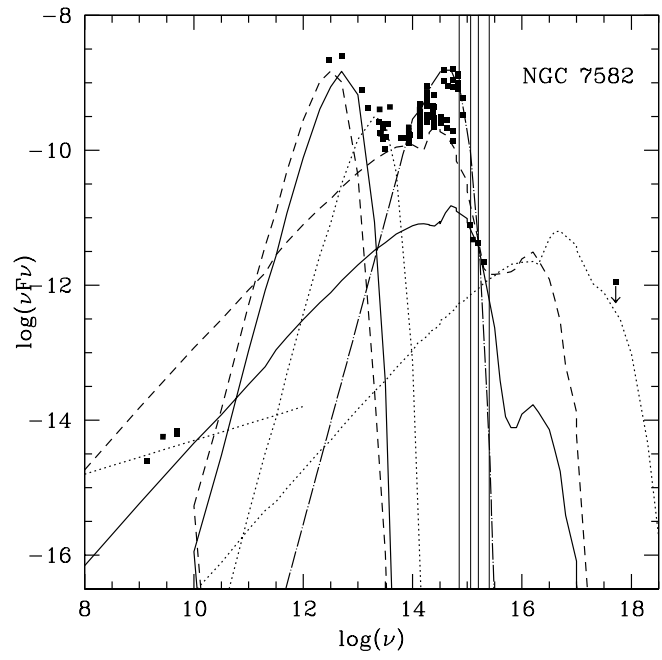
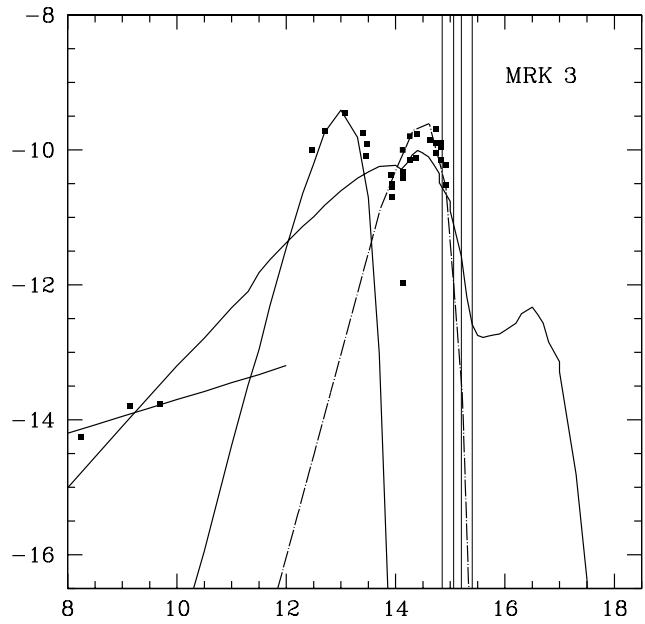
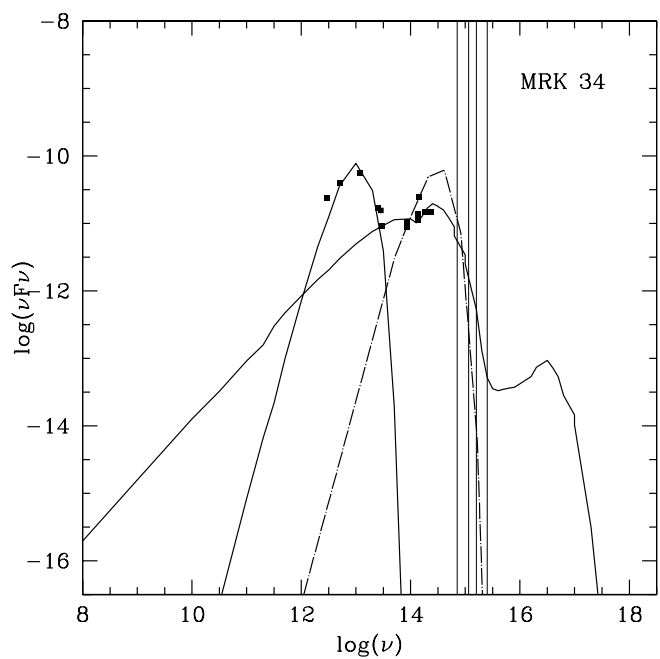
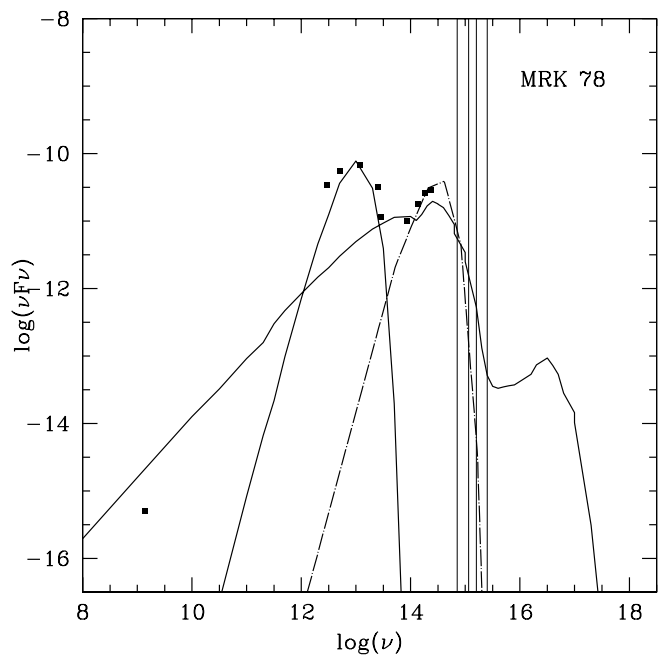
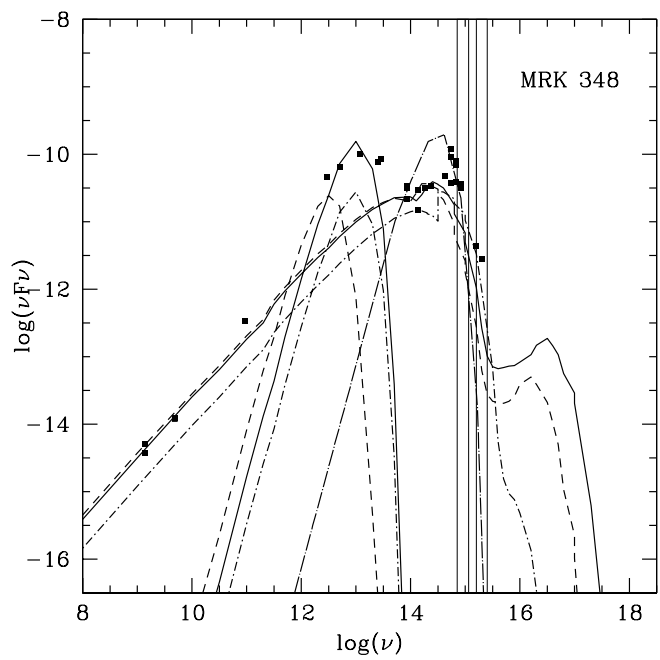
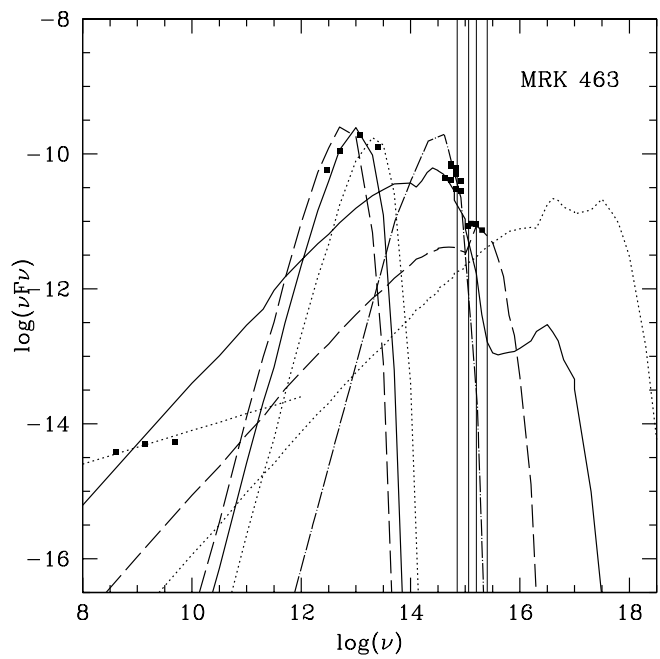
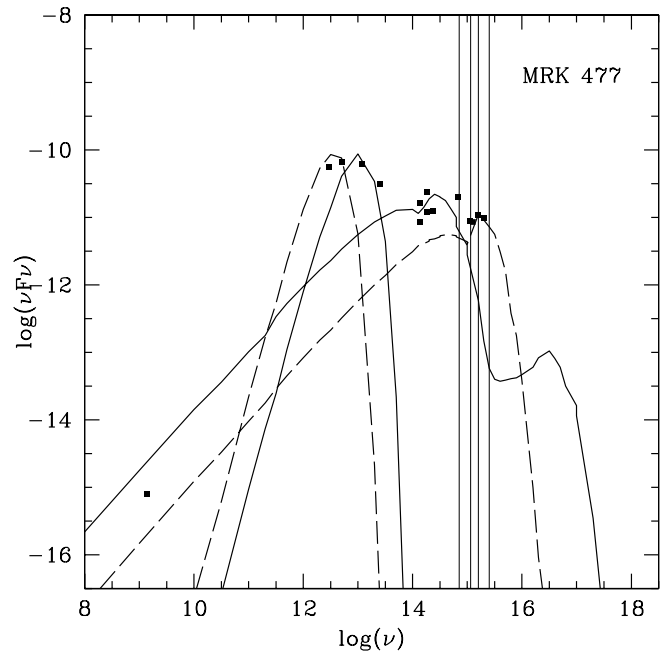
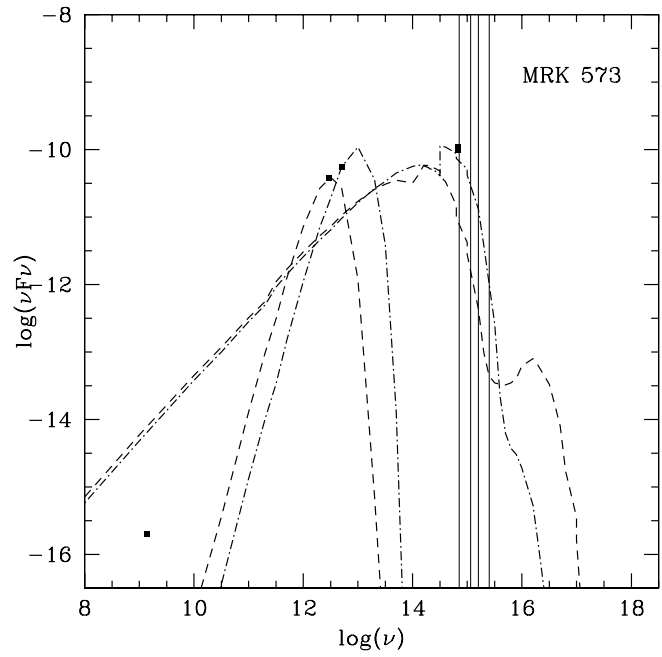
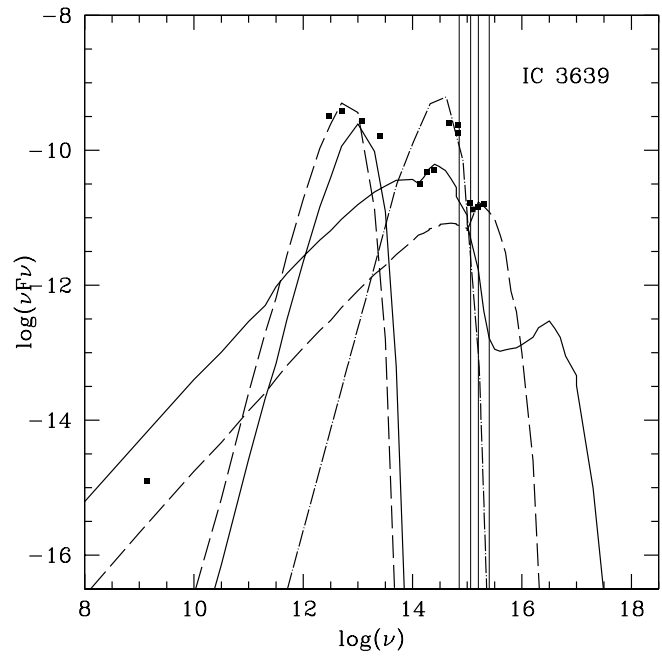
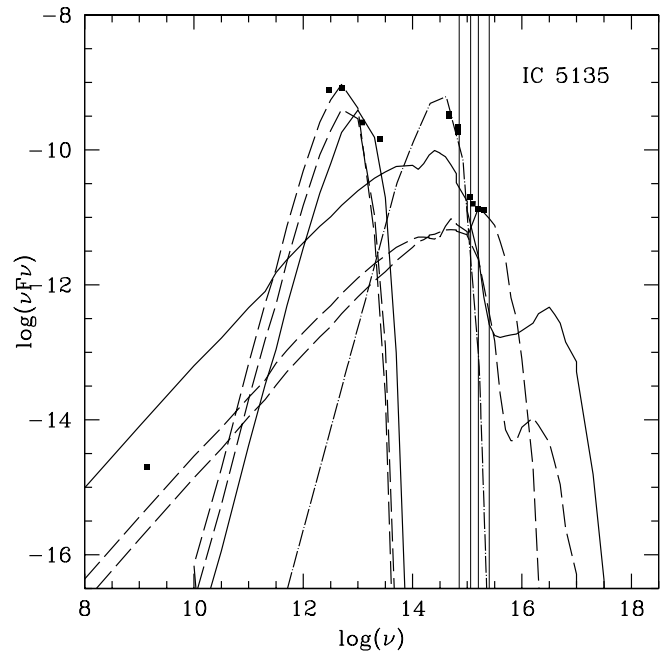
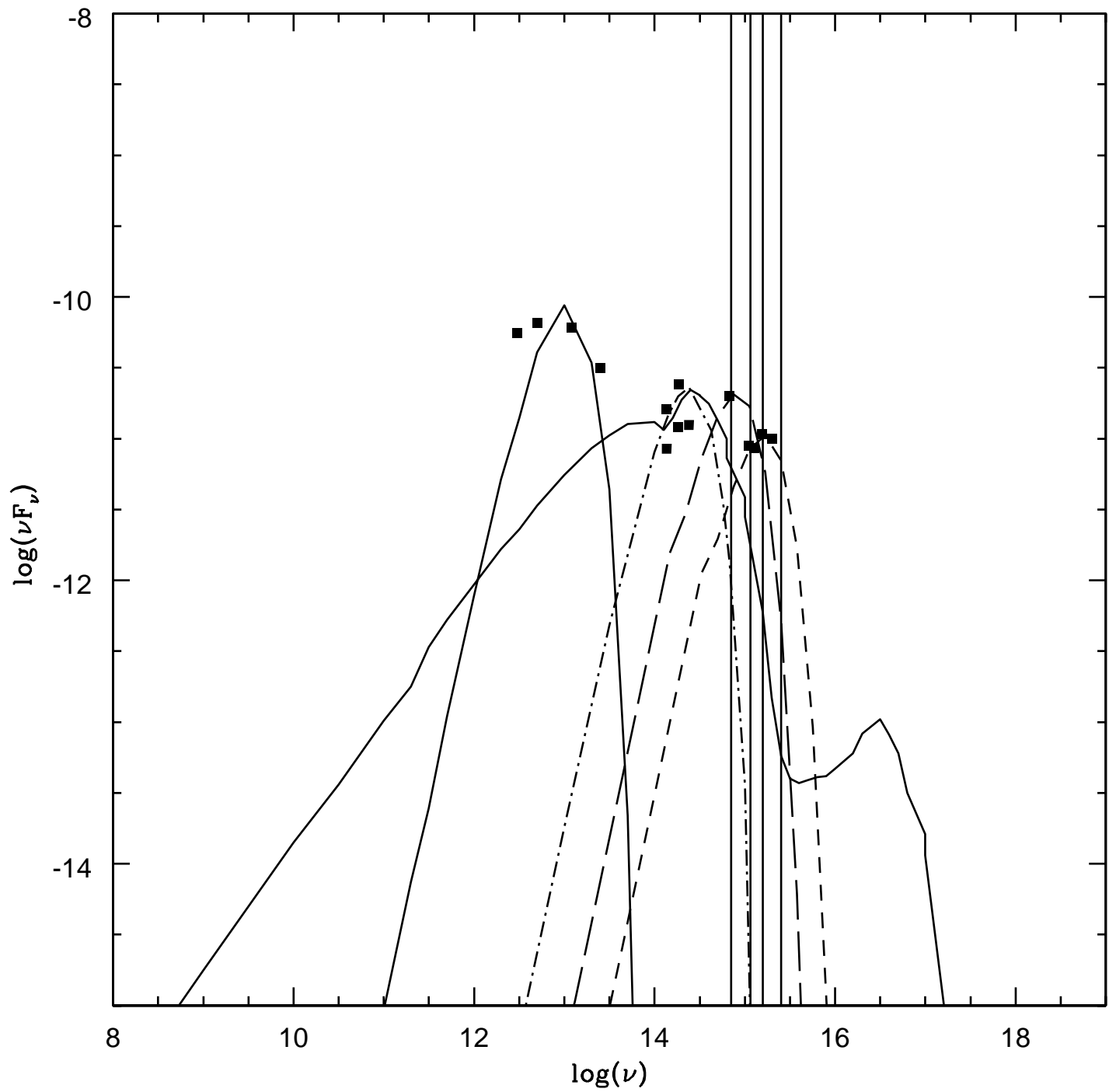
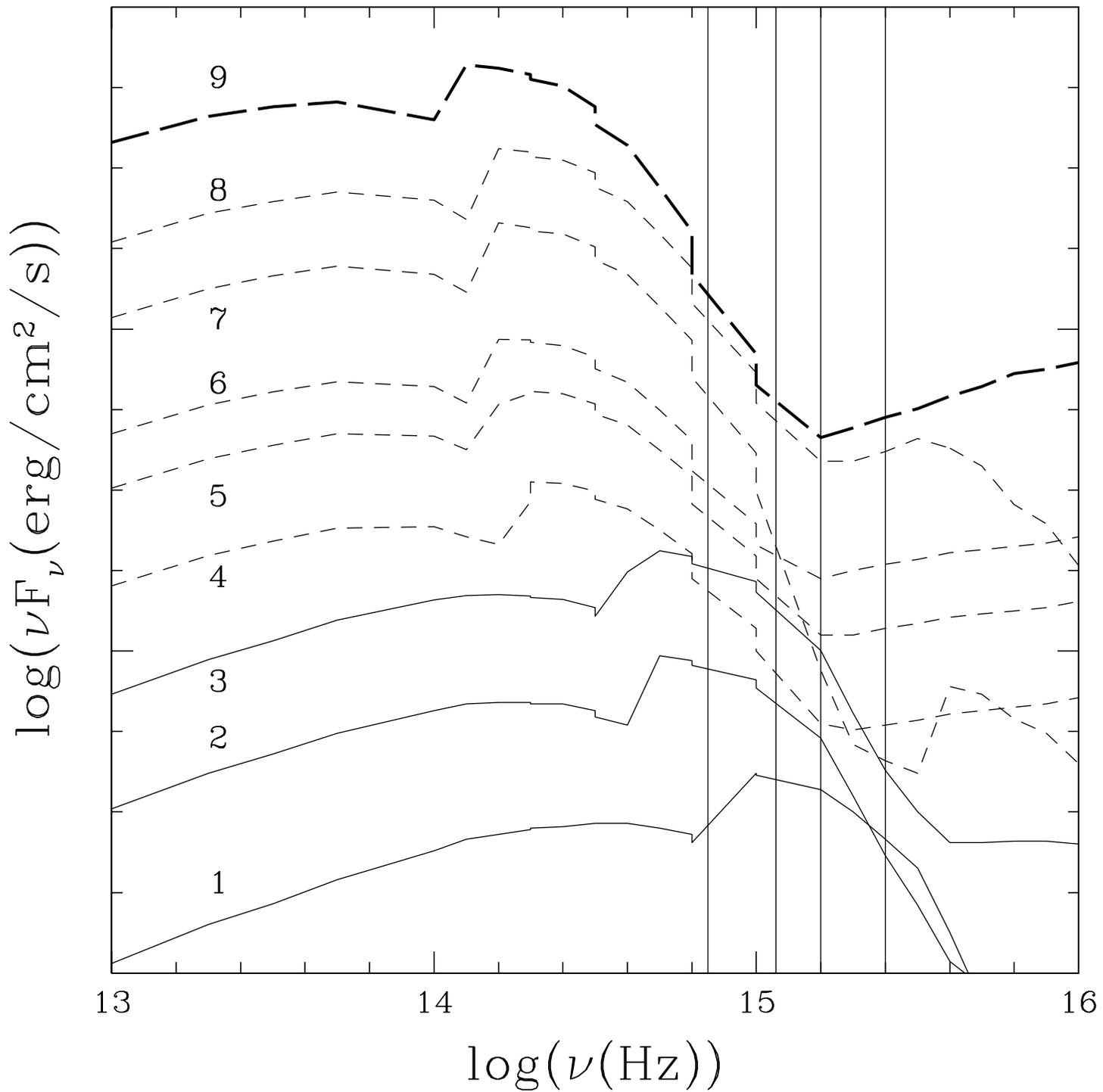
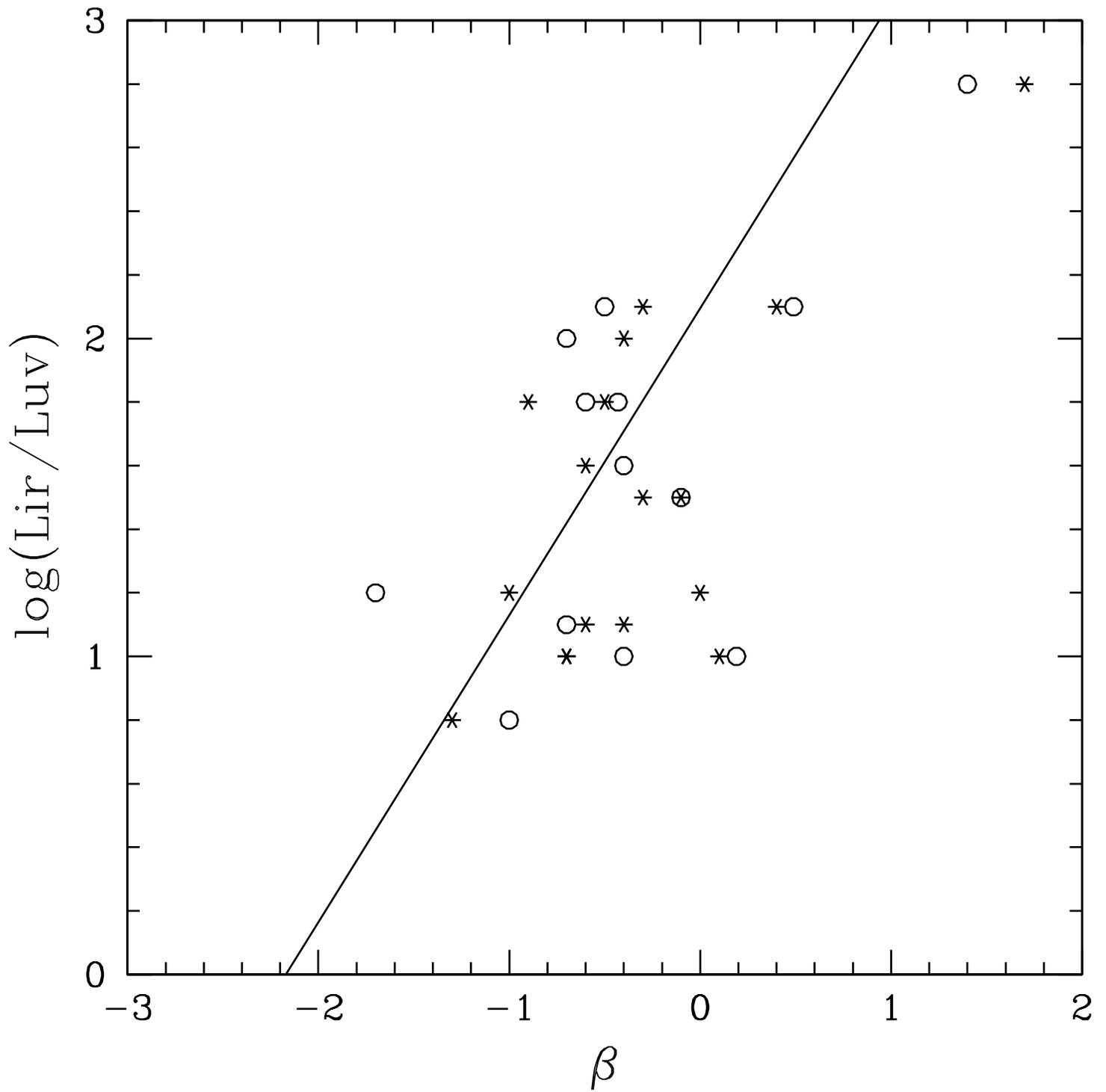
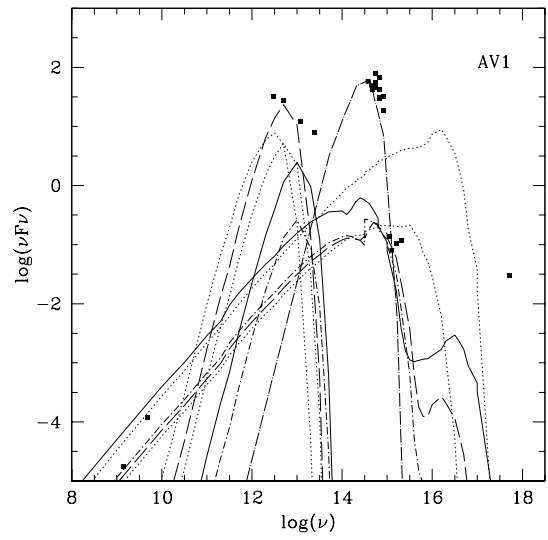
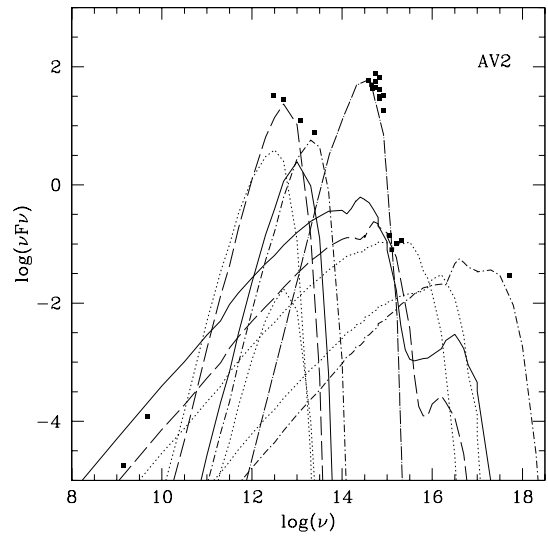
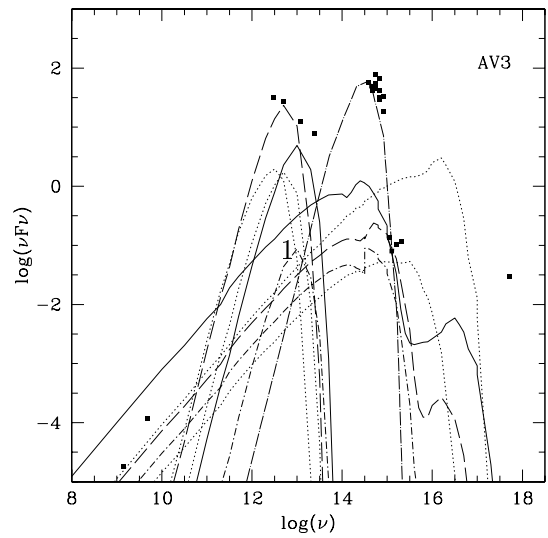
\end{document}